\documentclass[fleqn,10pt]{wlscirep}
\usepackage[utf8]{inputenc}
\usepackage[T1]{fontenc}
\usepackage{epstopdf}
\usepackage{xcolor}
\usepackage{amsmath,amssymb,graphicx}
\usepackage{blindtext}

\title{Quantum Zeno Repeaters}

\author[1*]{Veysel Bayrakci}
\author[2,1,3]{Fatih Ozaydin}
\affil[1]{Faculty of Engineering and Natural Sciences, Isik University, Sile, Istanbul, 34980, Turkey}
\affil[2]{Institute for International Strategy, Tokyo International University, 1-13-1 Matoba-kita, Kawagoe, Saitama 350-1197, Japan}
\affil[3]{CERN, 1211 Geneva 23, Switzerland}
\affil[*]{{\color{black}bayrakciveysel07@gmail.com }}

\begin{abstract}
Quantum repeaters pave the way for long-distance quantum communications and quantum Internet, and the idea of quantum repeaters is based on entanglement swapping which requires the implementation of controlled quantum gates.
Frequently measuring a quantum system affects its dynamics which is known as the quantum Zeno effect (QZE).
Beyond slowing down its evolution, QZE can be used to control the dynamics of a quantum system by introducing a carefully designed set of operations between measurements.
Here, we propose an entanglement swapping protocol based on QZE, which achieves almost unit fidelity. 
Implementation of our protocol requires only simple frequent threshold measurements and single particle rotations.
We extend the proposed entanglement swapping protocol to a series of repeater stations for constructing quantum Zeno repeaters which also achieve almost unit fidelity regardless of the number of repeaters.
Requiring no controlled gates, our proposal reduces the quantum circuit complexity of quantum repeaters. 
Our work has potential to contribute to long distance quantum communications and quantum computing via quantum Zeno effect.

\end{abstract}
\begin{document}
\flushbottom
\maketitle

\thispagestyle{empty}


\noindent
Due to inevitable attenuation in the channel, communication between two stations becomes impossible for long distances.
In classical communications, this problem is solved by repeaters based on simple signal amplification. 
However, because measuring the state of a quantum system alters its quantum state and due to no-cloning theorem~\cite{NC}, this idea is not applicable in quantum communications~\cite{cacciapuoti2019quantum,cacciapuoti2020entanglement}. 
The solution in quantum domain, i.e., the idea of quantum repeaters is based on the so-called entanglement swapping process~\cite{Azuma2015IEEE}.

\begin{figure}[h!]
	\centerline{\includegraphics[width=0.45\columnwidth]{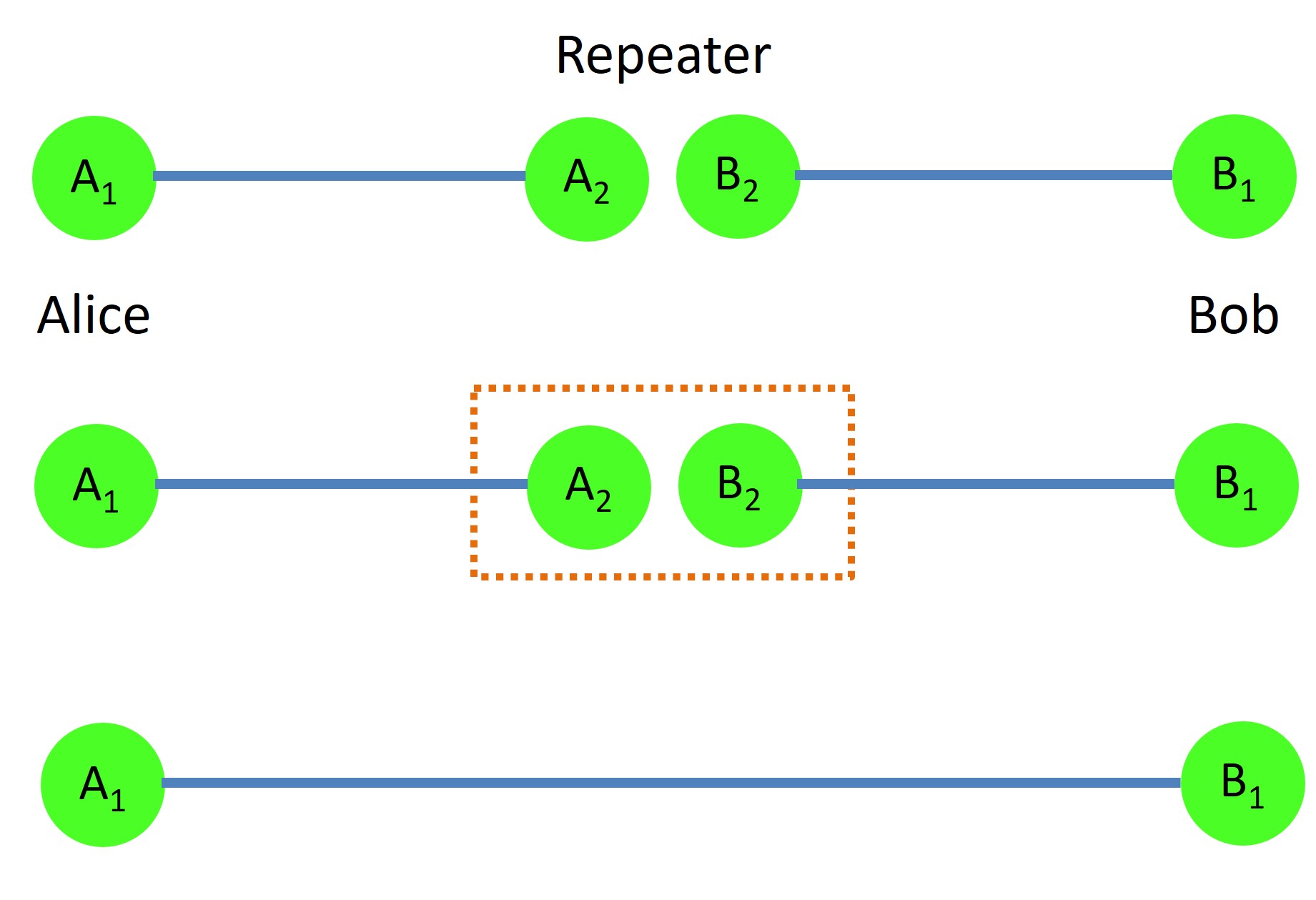}}
	\caption{Illustrating the entanglement swapping procedure. Possessing two qubits, $A_2$ entangled with Alice's qubit $A_1$, and $B_2$ entangled with Bob's qubit $B_1$, Repeater performs operations and measurements on $A_2$ and $B_2$, leaving Alice's and Bob's qubits entangled.}\label{fig:ShortRepeater}
\end{figure}

As illustrated in Fig.~\ref{fig:ShortRepeater}, the idea can be summarized as follows. 
Consider that the distance between two parties, Alice and Bob is beyond the limits of sharing entanglement reliably, and that the half of the distance is within the limits.
Placing a repeater station in the middle, Alice prepares a pair of entangled particles and sends one particle to the station.
Bob repeats the same procedure.
Then the repeater station applies local controlled-operations on the two particles it possesses, and the other two particles possessed by Alice and Bob become entangled.

In details, let a system of four qubits in the state $|\Psi_{A_1 A_2 B_2 B_1 }\rangle$ be initially shared among Alice, Repeater, and Bob; each qubit denoted as $A_1$, $A_2$ and $B_2$, and $B_1$, respectively, in the computational basis as

\begin{equation}\label{Intro:eq:initialstate}
	|\Psi_{A_1 A_2 B_2 B_1 }\rangle = \frac{|0_{A_1} 0_{A_2} \rangle + |1_{A_1} 1_{A_2} \rangle }{\sqrt{2}} \otimes \frac{|0_{B_2} 0_{B_1} \rangle + |1_{B_2} 1_{B_1} \rangle }{\sqrt{2}}.
\end{equation}

\noindent 
A controlled-NOT (CNOT) gate is applied to qubits $A_2$ and $B_2$ as the control and target qubits, respectively, followed by a Hadamard gate on $A_2$.
Then qubits $A_2$ and $B_2$ are measured in $z$-basis, yielding results $\{|i\rangle \}_{i=0,1}$.  
Measurement results are communicated through classical channels with Alice and Bob. 
Applying one of the single qubit operations $\{ I,\sigma_x,\sigma_z\}$ accordingly, Alice and Bob are left with and entangled pair of qubits,

\begin{equation}\label{Intro:eq:finalstate}
	|\Psi_{A_1 B_1 }\rangle = \frac{|0_{A_1} 0_{B_1} \rangle + |1_{A_1} 1_{B_1} \rangle }{\sqrt{2}},
\end{equation}

\noindent
where $I$ is the two-dimensional identity operator, and $\sigma_x$ and $\sigma_z$ are the Pauli operators.

\begin{figure}[h!]
	\centerline{\includegraphics[width=1\columnwidth]{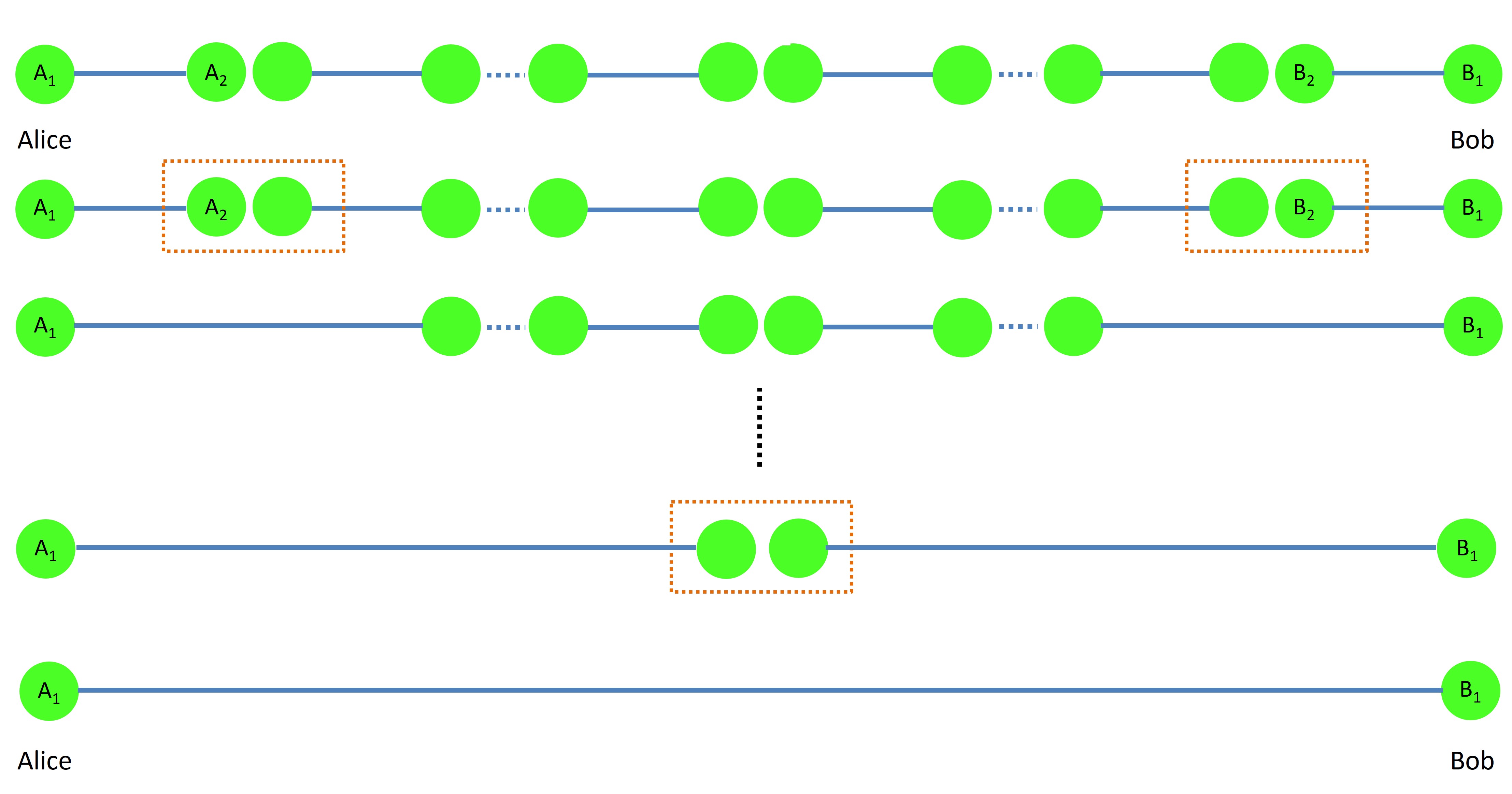}}
	\caption{Extending the entanglement swapping procedure in Fig.~\ref{fig:ShortRepeater} to long distances with many repeater stations in between.}\label{fig:LongRepeater}
\end{figure}

Extending the entanglement swapping process over a commensurate number of repeaters, Alice and Bob can share an entangled state, as shown in Fig.~\ref{fig:LongRepeater}, regardless of how long the distance is between them.
This makes the quantum repeaters essential for long distance quantum communication and quantum Internet, attracting an intense attention from both theoretical and experimental points of view. 

In addition to the photon loss, various types of noise pose a challenge.
Through a nested purification protocol Briegel et al. designed a quantum repeater mechanism to overcome the exponentially scaling of error probability with respect to the distance~\cite{briegel1998quantum}, and
enabling reliable communication despite the noise in the channel allows quantum key distribution (QKD) over long distances with unconditional security~\cite{doi:10.1126/science.283.5410.2050}.
Childress et al. considered active purification protocol at each repeater station for fault tolerant long distance quantum communication and proposed a physical realization of the protocol based on nitrogen-vacancy (NV) centers~\cite{childress2006fault}.
It was predicted that the hybrid design of van Loock et al. based on light-spin interaction can achieve 100 Hz over 1280 km with almost unit fidelity~\cite{PhysRevLett.96.240501}.
Generating encoded Bell pairs throughout the communication backbone, the protocol of Jiang et al. applies classical error correcting during simultaneous entanglement swapping and can extend the distance significantly~\cite{jiang2009quantum}.
Yang et al. have proposed a cavity-QED system which does not require the joint-measurement~\cite{PhysRevA.71.034312}, and showed that entanglement swapping can enable entanglement concentration for unknown entangled states~\cite{PhysRevA.71.044302}.

The light-matter interaction at repeater stations mainly for storing the quantum information in matter quantum memories was believed to be necessary, which makes the physical realization challenging.
However, designing an all-photonic quantum repeaters based on all flying qubits, Azuma et al. have shown that it is not the case, making a breakthrough in bringing the concept of quantum repeaters to reality~\cite{azuma2015all}.

Though requiring quantum memories at repeater stations, using spontaneous parametric downconversion sources, the nested purification~\cite{Chen2017} and fault-tolerant two-hierarchy entanglement swapping~\cite{PhysRevLett.119.170502} have been experimentally demonstrated.
Entangling electrons and nuclear spins through interactions with single photons, Kalb et al. have generated copies of remote entangled states towards quantum repeaters~\cite{doi:10.1126/science.aan0070}.
Recently, the idea of building quantum repeaters without quantum memory was experimentally demonstrated recently by Li et al. using linear optics~\cite{li2019experimental}. 
For a thorough review of recent advances in quantum repeaters, please refer to ref.~\cite{Yan_2021}.

Implementing the entanglement swapping procedure at each repeater station requires the realization of controlled two-qubit operations in the usual circuit model.
Regardless of the technology and type of physical particles used as qubits, realizing two-qubit logic operations is a bigger challenge than single-qubit operations.
Hence, in this work, we ask whether entanglement swapping can be implemented without controlled two-qubit operations, which could bring the quantum repeaters closer to reality. 
We consider quantum Zeno dynamics for serving this purpose.
Beyond practical concerns towards long distance quantum communication and quantum Internet, building quantum repeaters based on quantum Zeno dynamics have potential to contribute to fundamentals of quantum entanglement.

The quantum Zeno effect (QZE) can be described as follows~\cite{doi:10.1063/1.523304,Kofman2000}.
If a quantum system in state $|e\rangle$ initially (at $t=0$) evolves under Hamiltonian $\hat{H}$, 
the probability of finding it in the same state, i.e. the \textit{survival probability} at a later time (at $t>0$) is given as 
\begin{equation}
	p(t) = \left| \left\langle e \left|\exp \left( -{i\over \hslash} \hat{H} t\right) \right| e\right\rangle \right|^2.
\end{equation}
\noindent 
Assuming the Hamiltonian $\hat{H}$ with a finite variance $\langle V^2 \rangle$ and considering short times, the survival probability is found to be 
\begin{equation}
p(t) \approx 1- \frac{1}{\hslash^2} \langle V^2 \rangle t^2.
\end{equation}

\noindent
Now, let us assume ideal projective measurements on the system at intervals $\tau$. 
For $1/\tau \gg \langle V^2 \rangle ^{1 \over 2} / \hslash$, the survival probability is

\begin{equation}
	p^n(\tau) = p(t = n \tau) \approx \exp \left[ - \frac{1}{\hslash^2}( \langle V^2 \rangle \tau ) t \right].
\end{equation}

\noindent
In other words, the evolution of the system from the initial state slows down with $\tau$. 
What is more, for $\tau\rightarrow 0$, the survival probability $p(t)$ approaches $1$, which is widely considered as freezing the evolution of the system, such as in freezing the optical rogue waves~\cite{BAYINDIR2018141} and quantum chirps~\cite{BAYINDIR2021127096}.
It was also shown that the frequent measurements can be designed for accelerating the decay of the system, which is also known as the anti-Zeno effect~\cite{Kofman2000}.
Introducing a carefully designed set of quantum operations between measurements, QZE can be used to drive the a quantum system towards a target state, which is also known as the quantum Zeno dynamics (QZD).

One of the early experimental evidences of QZE was that in the the RF transition between two $^9\text{Be}^+$ ground state hyperfine levels, collapse to the initial state was observed if frequent measurements are performed~\cite{PhysRevA.41.2295}.
QZE has been studied for slowing down the system's evolution in Bose-Einstein condensates~\cite{Schäfer2014}, ion traps~\cite{PhysRevA.69.012303}, nuclear magnetic resonance~\cite{PhysRevA.87.032112}, cold atoms~\cite{PhysRevLett.87.040402}, cavity-QED~\cite{PhysRevLett.101.180402,PhysRevLett.105.213601,PhysRevA.86.032120} and large atomic systems~\cite{Signoles2014}.
QZE is being considered in various fundamental concepts. 
For example, it has been demonstrated in $PT$-symmetric systems in symmetric and broken-symmetric regions~\cite{Chen2021NPJ}.
Quantum heat engines have been attracting an increasing attention in quantum thermodynamics~\cite{tuncer2019work,dag2019temperature}, and Mukherjee et al. has recently discovered the advantages of anti-Zeno effect in fast-driven heat engines~\cite{Mukherjee2020}.

An interesting application of QZD in quantum information and computation is creating entanglement between two initially separated qubits by applying single-qubit operations and performing simple threshold measurements in an iterative way, without requiring a CNOT gate~\cite{PhysRevA.77.062339}.
Reducing the quantum circuit complexity by removing the controlled operations is promising for physical realizations. 
In a similar vein, recently, the activation of bound entanglement was shown to be enabled via QZD based on single particle rotations and threshold measurements~\cite{PhysRevA.105.022439}, which requires several three-level controlled operations, bound entangled states and classical communications otherwise in the original activation proposal by Horodecki et al.~\cite{PhysRevLett.82.1056}.
Quantum Zeno effect has been studied for generating multi-partite entanglement as well~\cite{CHEN2016140,doi:10.1126/science.aaa0754}, which is one of the most important problems attracting serious efforts in quantum science and technologies~\cite{DetExp,CavityRefs2,CavityRefs21,SRep_2020_10_3481,PRA_2021_103_052421}.

\section*{Results}
\begin{figure}[h!]
	\centerline{\includegraphics[width=1\columnwidth]{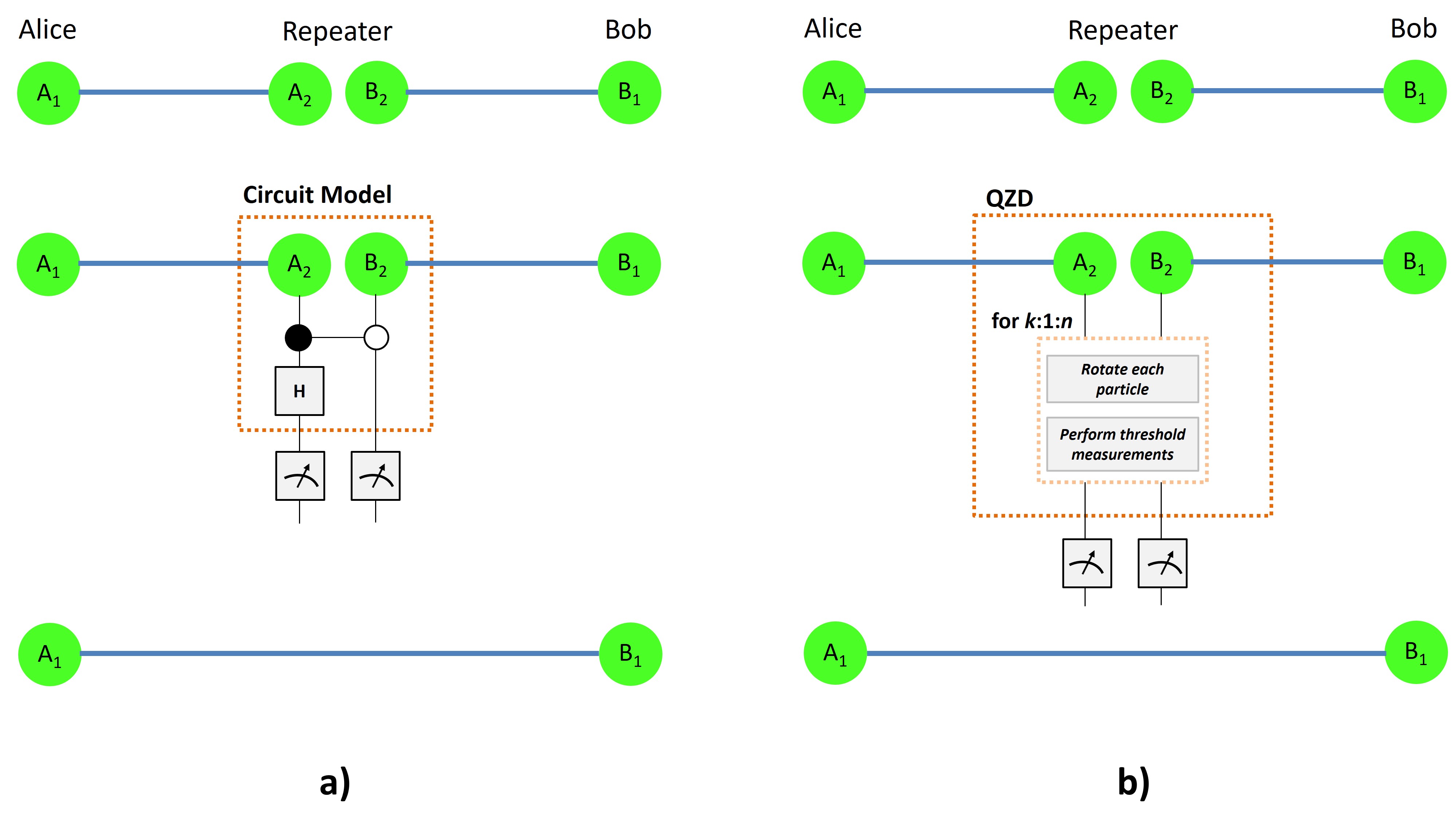}}
	\caption{Entanglement swapping procedure via \textbf{a)} the usual circuit model consisting of a CNOT and a Hadamard gate, \textbf{b)} the proposed QZD strategy consisting of only single qubit rotations and simple threshold measurements, requiring no controlled two-qubit gates.}\label{fig:CircuitVsZeno}
\end{figure}

Our QZD proposal for entanglement swapping (ES) starts with the joint system of two Bell states as in Eq.(\ref{Intro:eq:initialstate}), described by the density matrix $\rho_{A_1 A_2 B_2 B_1}$.
We set $\theta=\pi/180$ for simplicity, and through numerical studies we find the threshold measurement operators to be $J_1=|1\rangle\langle1|\otimes|1\rangle\langle1|$ with $J_0 = I-J_1$ in accordance with Ref.~\cite{PhysRevA.77.062339}.
First, let us examine the case of a single iteration, i.e., $n=1$ in the procedure illustrated in Fig.~\ref{fig:CircuitVsZeno}b.
After a single rotate-measure iteration on qubits $A_2$ and $B_2$, we proceed with the final measurement in $z-$basis. Finding $|0\rangle\otimes|0\rangle$, for example, the system of two qubits $A_1$ and $B_1$ is found approximately in the state
\begin{equation}
	\displaystyle \rho^1_{A_1 B_1} \!=\!\! 
	\left(
	\begin{array}{cccc}
		\ \ \ 0.9993 	& -0.0174 			& -0.0174 			& 0.0003  \\
	   -0.0174 			& \ \ \  0.0003 	& \ \ \  0.0003 	& 0  \\
	   -0.0174 			& \ \ \  0.0003 	& \ \ \  0.0003 	& 0  \\
		\ \ \ 0.0003 	&  0 		&  0 	& 0  \\
	\end{array}
	\right)
	\label{eq:rho_{A_1 B_1}},
\end{equation}

\noindent where the subscript denotes the number of iterations performed.

\begin{figure}
	\centering
	\includegraphics[width=0.7\columnwidth]{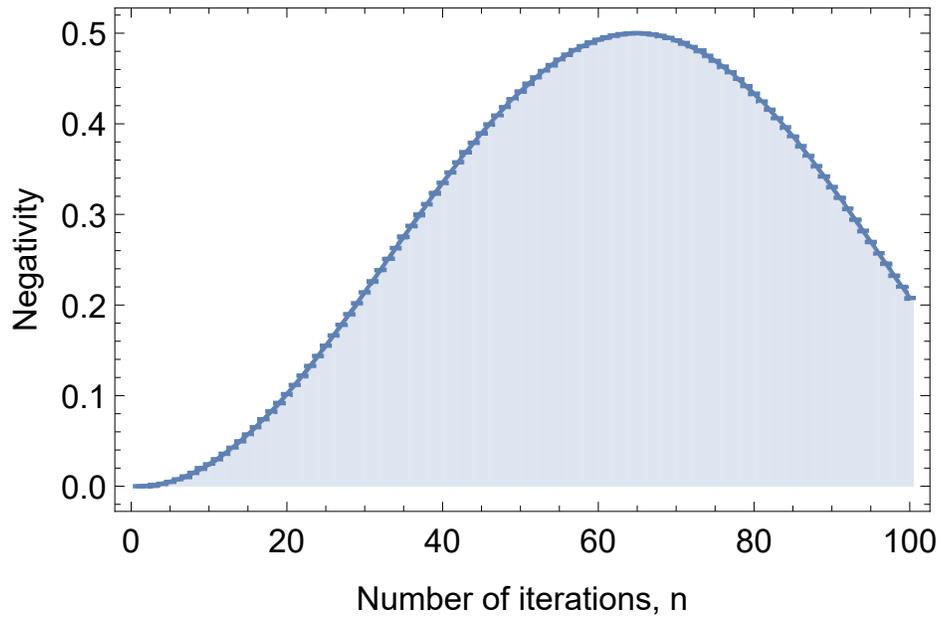}
	\caption{Negativity of the state obtained after $n$ rotate-measure iterations of QZD for realizing a single entanglement swapping as in Fig.~\ref{fig:CircuitVsZeno}b.}
	\label{fig:figplot}
\end{figure}

To find after how many iterations we should end the QZD procedure, we run the simulation one hundred times and end the procedure at $n$th run (consisting of $n$ iterations) with $n=1,2,...,100$.
After each, we calculate the negativity of the resulting state $\rho_{A_1 B_1}$, which we plot in Fig.~\ref{fig:figplot}.
Our simulation shows that within this setting, the resulting state after $n=65$ iterations (and after a $\sigma_x$ by Alice following the $z$-basis measurement result) is approximately
\begin{equation}
	\displaystyle \rho^{65}_{A_1 B_1} \!=\!\! 
	\left(
	\begin{array}{cccc}
		\ \ \ 0.4993 	& \ \ \    -0.0174  & \ \ \  0.0193 	& \ \ \ 0.4993  \\
		     -0.0174	& \ \ \ 	0.0006  &       -0.0006 	& -0.0174  \\
		\ \ \ 0.0193	&          -0.0006 	& \ \ \  0.0007 	& \ \ \ 0.0193  \\
		 \ \ \    0.4993 	&  		   -0.0174		&  \ \ \ 0.0193 	& \ \ \ 0.4993   \\
	\end{array}
	\right)
	\label{eq:rho65_{A_1 B_1}},
\end{equation}

\noindent with negativity $N(\rho^{65}_{A_1 B_1})=0.4999$.
The fidelity of this state to the maximally entangled Bell state in Eq.(\ref{Intro:eq:finalstate}) is found to be 
$F(\rho^{65}_{A_1 B_1}, |\Psi_{A_1 B_1 }\rangle) := \langle \Psi_{A_1 B_1}| \ \rho^{65}_{A_1 B_1} \ |\Psi_{A_1 B_1} \rangle=0.9986$.
This result shows that the entanglement swapping can be implemented with almost a unit fidelity by QZD, i.e., only through single qubit rotations and simple threshold measurements, without requiring any controlled operations, reducing the complexity of quantum repeaters significantly in terms of controlled two-qubit operations.

\begin{figure}
	\centering
	\includegraphics[width=0.7\columnwidth]{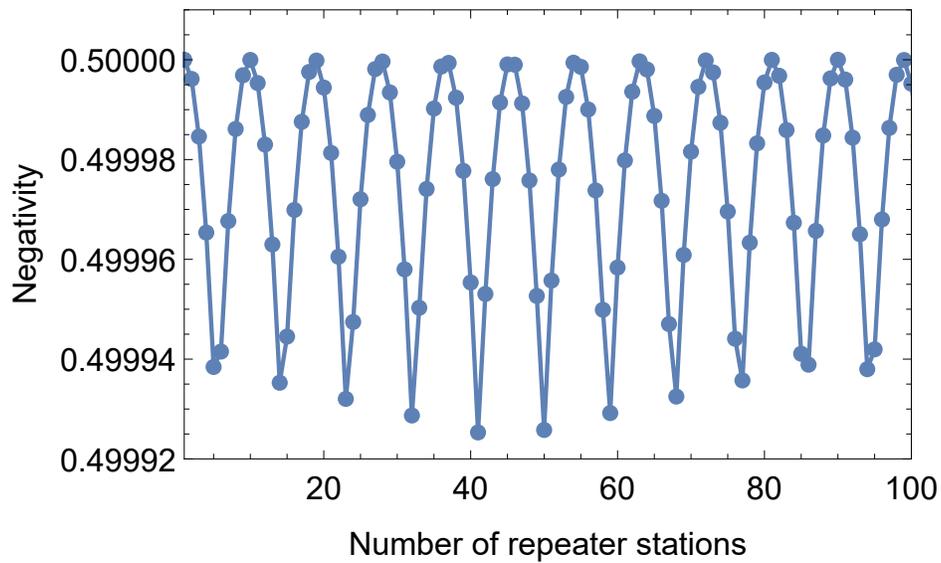}
	\caption{Negativity of the obtained state after ES over a hundred repeater stations.}
	\label{fig:PlotRepeater}
\end{figure}

Next, we extend the QZD-based ES to a series of repeater stations.
We consider the state $\rho^{65}_{A_1 B_1}$ obtained from the first ES to be one of the two states of the second ES and the other being a maximally entangled state equivalent to $|\Psi_{A_1 B_1}\rangle$. 
The obtained non-maximally entangled two-qubit state in the second ES will then be considered for the third ES with a maximally entangled state, and so on for enabling long-distance quantum communication via quantum Zeno repeaters (QZR).
At the first glance, it might be expected to obtain the new state with decaying negativity at each ES, vanishing with increasing distance, i.e., the number of repeater stations.
However, this is not the case, demonstrating the strength of our proposed QZD.
The negativity of the state obtained from each ES exhibits an oscillation. For example, after it decreases to $0.499938$ in the fifth ES, it increases to $0.499942$ in the sixth.
We plot the negativity values of the states obtained over $100$ repeater stations in Fig.~\ref{fig:PlotRepeater}.
To provide a clearer presentation of the turning point of the negativity, we also plot the results for the first $9$ states in Fig.~\ref{fig:PlotRepeaterFirst10}, and provide the corresponding density matrices in the Appendix.

\section*{Discussion}
A major contribution of the proposed quantum Zeno repeaters (QZR) is to reduce the quantum circuit complexity of repeaters in terms of controlled multi-particle operations as illustrated in Fig.~\ref{fig:CircuitVsZeno}a, which is more challenging than single  particle operations in any technology in principle.
Because our QZR protocol can be extended to multi-level particles, this reduction would be even more significant than the case of qubits.
However, beyond practical concerns for reducing the quantum circuit complexity, we believe showing that quantum repeaters can be realized via quantum Zeno dynamics contributes to our understanding of quantum entanglement and measurements.

One of the drawbacks of our protocol is that under ideal conditions except for the attenuation in the channel which requires the repeaters in the first place, not exactly but almost unit fidelity can be achieved. 
However, over $0.998$ fidelity can be tolerated in physical realizations especially given that the fidelities will decrease in both approaches.

A more serious drawback could be the increased latency. Repeaters based on the standard circuit model requires the implementation of only two logic operations -though one being the controlled multi-particle operation. 
Our protocol requires the implementation of several single-particle operations and simple threshold measurements, instead. 
This would take a longer time depending on the physical realization, introducing a higher latency, which might not be desired especially considering on-demand systems and designs without quantum memory.

The slight increase in the negativity does not violate the monotonicity of entanglement measure since a single entangled state with negativity $\approx 0.5$ is obtained out of two entangled states with total negativity $\approx 0.5$.
The reason we prefer the negativity entanglement measure as the key performance indicator over the fidelity is as follows. 
In each ES, the resulting state is close to one of the four Bell states, which are equivalent under local operations and classical communications~\cite{NC}.
Hence, rather than finding which Bell state it is the closest to and then calculating the fidelity each time, for simplicity, we chose to calculate the negativity which is invariant under Pauli operators that the parties can apply to convert one Bell state to another.

Note that while our QZD-based ES protocol requires $65$ iterations in the first repeater, next  repeaters might require a different number of iterations. 
Our simulation picks the best number for each repeater station and the presented results are based on the the best outcomes.

For the physical realization of our QZD protocol, we consider the superconducting circuit proposed by Wang et al.~\cite{PhysRevA.77.062339} where the threshold measurements can be implemented by Josephson-junction circuit with flux qubits.
In the same work, physical imperfections were also analyzed by considering a possible deviation from the rotation angle $\theta$ in each iteration. 
It was found that in the case of several iterations, the impact of the deviations is minimized, implying the robustness of the protocol.
Because our protocol follows a similar rotate-measure procedure with many iterations, we consider a similar inherent robustness, too.

Apart from the attenuation in the channel, we have studied our protocol under ideal conditions. 
However, because QZE has been mostly considered for protecting the system from noise induced by interactions with the environment, it would be interesting as a future research to design a QZD protocol with an inherent error-correction mechanism.

\section*{Methods}

In each iteration of QZD, a set of two basic operations are performed. 
First, the following rotation operation is applied on each of the two qubits at the repeater station,
\begin{equation}\label{Methods:eq:singlerotation}
	\displaystyle R(\theta) = 
	\left(
	\begin{array}{cc}
		\cos \theta & - \sin \theta  \\
		\sin \theta & \ \   \  \cos \theta  \\
	\end{array}
	\right),
\end{equation}

\noindent which is followed by the threshold measurements on each qubit in concern, defined by the measurement operators
\begin{equation} \label{Methods:eq:Js}
	J_1= |i\rangle\langle i| \otimes |j\rangle\langle j|, \ \ \ \ \ J_0 = I^{\otimes 2} -J_1
\end{equation}

\noindent with $i, j \in {0, 1}$ and $I$ being the two-dimensional identity operator.
The system will be found in $|i\rangle|j\rangle$ state with a small probability $\epsilon$, and with $1-\epsilon$ probability it is projected to the $J_0$ subspace. 
Hence, in each iteration, the state of the system evolves in the rotate-measure procedure as $\rho \rightarrow \rho^{r} \rightarrow \rho^{rm}$ where 
\begin{equation}\label{Methods:eq:rotate}
	\rho^{r} = (I \otimes R(\theta)^{\otimes 2} \otimes I ) \ \rho \ (I \otimes R(\theta)^{\otimes 2} \otimes I )^{\dagger},
\end{equation}
\noindent and
\begin{equation} \label{Methods:eq:measure}
	\rho^{rm} = {(I \otimes J_0 \otimes I) \ \rho^{r} \ (I \otimes J_0 \otimes I)^{\dagger} \over \text{Tr}[(I \otimes J_0 \otimes I) \ \rho^{r} \ (I \otimes J_0 \otimes I)^{\dagger}]}.
\end{equation}

After $n$ iterations, the QZD process is over and similar to the circuit model computation, two qubits at the repeater are measured in $z-$basis, and according to the results of this final measurement communicated over a classical channel, one of the Pauli operators $\{I,\sigma_x, \sigma_z\}$ is applied to the qubits of Alice and Bob, leaving them not exactly in the Bell state but in the state $\rho '$ with almost a unit fidelity to a Bell state.
Here, $\{i, j \}$ of $J_1$, the rotation angle $\theta$ and the number of iterations $n$ are to be determined by numerical simulations for achieving the closest $\rho '$ to a maximally entangled state.
Note that in each iteration for each qubit, considering a different $\theta$ could improve the performance with the drawback expanding the search space significantly. 
For simplicity, we fix $\theta$ for both qubits throughout the process. 

For extending the above entanglement swapping procedure to a series of repeater stations, we can assume that the entanglement swapping (ES) starts from both ends and continues towards the repeater station in the middle as in Fig.\ref{fig:LongRepeater}.
Therefore, although assuming that the first ES starts with maximally entangled states, the non-maximally entangled state $\rho '$ is obtained which is to be used in the next ES, creating $\rho ''$ state with a smaller fidelity to the maximally entangled state than $\rho '$.
Our numerical simulation takes into account the generated non-maximally entangled state being the output of each ES as the input to the next ES.

The negativity of a two-qubit state $\rho$ is found by the absolute sum of its negative eigenvalues $\mu_i$ of after partial transposition $\rho^{\Gamma_A}$ with respect to subsystem $A$ as 
\begin{equation}
	N(\rho) \equiv { ||\rho^{\Gamma_A}||_1 - 1 \over 2},
\end{equation}
\noindent
where $||A||_1$ is the trace norm of the operator $A$~\cite{PhysRevA.65.032314}.

\section*{Appendix}

In the Appendix, we provide the density matrices of the states obtained through ES over nine repeater stations, where $\rho_j$ denotes the state after $j$th ES.

\begin{equation}
	\displaystyle \rho_{1}  =  
	\left(
	\begin{array}{cccc}
		0.000753558	& 0.0193976  	& 0.0193976 	& -0.000677401  \\
		0.0193976	& 0.499319  	& 0.499319 	& -0.0174372  \\
		0.0193976	& 0.499319  	& 0.499319 	& -0.0174372  \\
		-0.000677401 	& -0.0174372		& -0.0174372		& 0.000608941   \\
	\end{array}
	\right),
\end{equation}

\begin{equation}
	\displaystyle \rho_{2} =  
	\left(
	\begin{array}{cccc}
		0.500137	& 0.00196063  	& 0.00196063 	& 0.499992  \\
		0.00196063	& 7.68598\text{x}10^{-6}  	& 7.68601\text{x}10^{-6} 	& 0.00196006  \\
		0.00196063	& 7.68601\text{x}10^{-6}		& 7.68601\text{x}10^{-6} 	& 0.00196006  \\
		0.499992 	& 0.00196006		& 0.00196006		& 0.499848   \\
	\end{array}
	\right),
\end{equation}

\begin{equation}
	\displaystyle \rho_{3}  =  
	\left(
	\begin{array}{cccc}
		0.000913537		& 0.0213573  	& 0.0213573 	& -0.000661988  \\
		0.0213573		& 0.499303  	& 0.499303 	& -0.0154764  \\
		0.0213573		& 0.499303		& 0.499303 	& -0.0154764  \\
		-0.000661988 	& -0.0154764		& -0.0154764		& 0.000479705   \\
	\end{array}
	\right),
\end{equation}

\begin{equation}
	\displaystyle \rho_{4} =  
	\left(
	\begin{array}{cccc}
		0.500258	& 0.00392158  		& 0.00392158 	& 0.499969  \\
		0.00392158	& 0.0000307416  	& 0.0000307417 	& 0.00391931  \\
		0.00392158	& 0.0000307417  	& 0.0000307417 	& 0.00391931  \\
		0.499969 	& 0.00391931		& 0.00391931	& 0.49968   \\
	\end{array}
	\right),
\end{equation}

\begin{equation}
	\displaystyle \rho_{5}  =  
	\left(
	\begin{array}{cccc}
		0.0000692054	& 0.00588032  	& 0.00588193 	& -0.000159162  \\
		0.00588032		& 0.499645  	& 0.499782 	& -0.0135238  \\
		0.00588193		& 0.499782		& 0.499919 	& -0.0135275  \\
		-0.000159162 	& -0.0135238	& -0.0135275		& 0.000366047   \\
	\end{array}
	\right),
\end{equation}

\begin{equation}
	\displaystyle \rho_{6} =  
	\left(
	\begin{array}{cccc}
		0.499725	& 0.00588306  	& -0.0115664 	& 0.499832  \\
		0.00588306	& 0.0000692589  	& -0.000136167 	& 0.00588432  \\
		-0.0115664	& -0.000136167		& 0.000267713 	& -0.0115689  \\
		0.499832 	& 0.00588432		& -0.0115689		& 0.499938   \\
	\end{array}
	\right),
\end{equation}

\begin{equation}
	\displaystyle \rho_{7}  =  
	\left(
	\begin{array}{cccc}
		0.000123053	& 0.00784181  	& 0.00784289 	& -0.00018147  \\
		0.00784181	& 0.499736  	& 0.499805 	& -0.0115645  \\
		0.00784289	& 0.499805		& 0.499873 	& -0.0115661  \\
		-0.00018147 	& -0.0115645		& -0.0115661		& 0.000267619   \\
	\end{array}
	\right),
\end{equation}

\begin{equation}
	\displaystyle \rho_{8} =  
	\left(
	\begin{array}{cccc}
		0.499815	& 0.00784462  	& -0.0096067 	& 0.499846  \\
		0.00784462	& 0.000123122  	& -0.000150778 	& 0.00784511  \\
		-0.0096067	& -0.000150778		& 0.000184646 	& -0.00960729  \\
		0.499846 	& 0.00784511		& -0.00960729		& 0.499877   \\
	\end{array}
	\right),
\end{equation}

\begin{equation}
	\displaystyle \rho_{9}  =  
	\left(
	\begin{array}{cccc}
		0.00019229	& 0.0098035  	& 0.0098035 	& -0.000188386  \\
		0.0098035	& 0.499812  	& 0.499812 	& -0.00960447  \\
		0.0098035	& 0.499812		& 0.499812 	& -0.00960447  \\
		-0.000188386 	& -0.00960447		& -0.00960447		& 0.000184561   \\
	\end{array}
	\right).
\end{equation}

\section*{Availability of Data and Materials}

All data generated or analysed during this study are included in this published article.

\bibliography{Manuscript-ZenoRepeater}

\begin{thebibliography}{10}
\urlstyle{rm}
\expandafter\ifx\csname url\endcsname\relax
  \def\url#1{\texttt{#1}}\fi
\expandafter\ifx\csname urlprefix\endcsname\relax\def\urlprefix{URL }\fi
\expandafter\ifx\csname doiprefix\endcsname\relax\def\doiprefix{DOI: }\fi
\providecommand{\bibinfo}[2]{#2}
\providecommand{\eprint}[2][]{\url{#2}}

\bibitem{NC}
\bibinfo{author}{Nielsen, M.~A.} \& \bibinfo{author}{Chuang, I.~L.}
\newblock \emph{\bibinfo{title}{Quantum Computation and Quantum Information:
  10th Anniversary Edition}} (\bibinfo{publisher}{Cambridge University Press},
  \bibinfo{year}{2011}).

\bibitem{cacciapuoti2019quantum}
\bibinfo{author}{Cacciapuoti, A.~S.} \emph{et~al.}
\newblock \bibinfo{journal}{\bibinfo{title}{Quantum internet: networking
  challenges in distributed quantum computing}}.
\newblock {\emph{\JournalTitle{IEEE Network}}} \textbf{\bibinfo{volume}{34}},
  \bibinfo{pages}{137--143},
  \doiprefix\url{https://doi.org/10.1109/MNET.001.1900092}
  (\bibinfo{year}{2019}).

\bibitem{cacciapuoti2020entanglement}
\bibinfo{author}{Cacciapuoti, A.~S.}, \bibinfo{author}{Caleffi, M.},
  \bibinfo{author}{Van~Meter, R.} \& \bibinfo{author}{Hanzo, L.}
\newblock \bibinfo{journal}{\bibinfo{title}{When entanglement meets classical
  communications: Quantum teleportation for the quantum internet}}.
\newblock {\emph{\JournalTitle{IEEE Transactions on Communications}}}
  \textbf{\bibinfo{volume}{68}}, \bibinfo{pages}{3808--3833},
  \doiprefix\url{https://doi.org/10.1109/TCOMM.2020.2978071}
  (\bibinfo{year}{2020}).

\bibitem{Azuma2015IEEE}
\bibinfo{author}{Munro, W.~J.}, \bibinfo{author}{Azuma, K.},
  \bibinfo{author}{Tamaki, K.} \& \bibinfo{author}{Nemoto, K.}
\newblock \bibinfo{journal}{\bibinfo{title}{Inside quantum repeaters}}.
\newblock {\emph{\JournalTitle{IEEE Journal of Selected Topics in Quantum
  Electronics}}} \textbf{\bibinfo{volume}{21}}, \bibinfo{pages}{78--90},
  \doiprefix\url{https://doi.org/10.1109/JSTQE.2015.2392076}
  (\bibinfo{year}{2015}).

\bibitem{briegel1998quantum}
\bibinfo{author}{Briegel, H.-J.}, \bibinfo{author}{D{\"u}r, W.},
  \bibinfo{author}{Cirac, J.~I.} \& \bibinfo{author}{Zoller, P.}
\newblock \bibinfo{journal}{\bibinfo{title}{Quantum repeaters: the role of
  imperfect local operations in quantum communication}}.
\newblock {\emph{\JournalTitle{Phys. Rev. Lett.}}}
  \textbf{\bibinfo{volume}{81}}, \bibinfo{pages}{5932},
  \doiprefix\url{https://doi.org/10.1103/PhysRevLett.81.5932}
  (\bibinfo{year}{1998}).

\bibitem{doi:10.1126/science.283.5410.2050}
\bibinfo{author}{Lo, H.-K.} \& \bibinfo{author}{Chau, H.~F.}
\newblock \bibinfo{journal}{\bibinfo{title}{Unconditional security of quantum
  key distribution over arbitrarily long distances}}.
\newblock {\emph{\JournalTitle{Science}}} \textbf{\bibinfo{volume}{283}},
  \bibinfo{pages}{2050--2056},
  \doiprefix\url{https://doi.org/10.1126/science.283.5410.2050}
  (\bibinfo{year}{1999}).

\bibitem{childress2006fault}
\bibinfo{author}{Childress, L.}, \bibinfo{author}{Taylor, J.},
  \bibinfo{author}{S{\o}rensen, A.~S.} \& \bibinfo{author}{Lukin, M.}
\newblock \bibinfo{journal}{\bibinfo{title}{Fault-tolerant quantum
  communication based on solid-state photon emitters}}.
\newblock {\emph{\JournalTitle{Phys. Rev. Lett.}}}
  \textbf{\bibinfo{volume}{96}}, \bibinfo{pages}{070504},
  \doiprefix\url{https://doi.org/10.1103/PhysRevLett.96.070504}
  (\bibinfo{year}{2006}).

\bibitem{PhysRevLett.96.240501}
\bibinfo{author}{van Loock, P.} \emph{et~al.}
\newblock \bibinfo{journal}{\bibinfo{title}{Hybrid quantum repeater using
  bright coherent light}}.
\newblock {\emph{\JournalTitle{Phys. Rev. Lett.}}}
  \textbf{\bibinfo{volume}{96}}, \bibinfo{pages}{240501},
  \doiprefix\url{https://doi.org/10.1103/PhysRevLett.96.240501}
  (\bibinfo{year}{2006}).

\bibitem{jiang2009quantum}
\bibinfo{author}{Jiang, L.} \emph{et~al.}
\newblock \bibinfo{journal}{\bibinfo{title}{Quantum repeater with encoding}}.
\newblock {\emph{\JournalTitle{Physical Review A}}}
  \textbf{\bibinfo{volume}{79}}, \bibinfo{pages}{032325},
  \doiprefix\url{https://doi.org/10.1103/PhysRevA.79.032325}
  (\bibinfo{year}{2009}).

\bibitem{PhysRevA.71.034312}
\bibinfo{author}{Yang, M.}, \bibinfo{author}{Song, W.} \& \bibinfo{author}{Cao,
  Z.-L.}
\newblock \bibinfo{journal}{\bibinfo{title}{Entanglement swapping without joint
  measurement}}.
\newblock {\emph{\JournalTitle{Phys. Rev. A}}} \textbf{\bibinfo{volume}{71}},
  \bibinfo{pages}{034312},
  \doiprefix\url{https://doi.org/10.1103/PhysRevA.71.034312}
  (\bibinfo{year}{2005}).

\bibitem{PhysRevA.71.044302}
\bibinfo{author}{Yang, M.}, \bibinfo{author}{Zhao, Y.}, \bibinfo{author}{Song,
  W.} \& \bibinfo{author}{Cao, Z.-L.}
\newblock \bibinfo{journal}{\bibinfo{title}{Entanglement concentration for
  unknown atomic entangled states via entanglement swapping}}.
\newblock {\emph{\JournalTitle{Phys. Rev. A}}} \textbf{\bibinfo{volume}{71}},
  \bibinfo{pages}{044302},
  \doiprefix\url{https://doi.org/10.1103/PhysRevA.71.044302}
  (\bibinfo{year}{2005}).

\bibitem{azuma2015all}
\bibinfo{author}{Azuma, K.}, \bibinfo{author}{Tamaki, K.} \&
  \bibinfo{author}{Lo, H.-K.}
\newblock \bibinfo{journal}{\bibinfo{title}{All-photonic quantum repeaters}}.
\newblock {\emph{\JournalTitle{Nature Communications}}}
  \textbf{\bibinfo{volume}{6}}, \bibinfo{pages}{1--7} (\bibinfo{year}{2015}).

\bibitem{Chen2017}
\bibinfo{author}{Chen, L.-K.} \emph{et~al.}
\newblock \bibinfo{journal}{\bibinfo{title}{Experimental nested purification
  for a linear optical quantum repeater}}.
\newblock {\emph{\JournalTitle{Nature Photonics}}}
  \textbf{\bibinfo{volume}{11}}, \bibinfo{pages}{695--699},
  \doiprefix\url{https://doi.org/10.1038/s41566-017-0010-6}
  (\bibinfo{year}{2017}).

\bibitem{PhysRevLett.119.170502}
\bibinfo{author}{Xu, P.} \emph{et~al.}
\newblock \bibinfo{journal}{\bibinfo{title}{Two-hierarchy entanglement swapping
  for a linear optical quantum repeater}}.
\newblock {\emph{\JournalTitle{Phys. Rev. Lett.}}}
  \textbf{\bibinfo{volume}{119}}, \bibinfo{pages}{170502},
  \doiprefix\url{https://doi.org/10.1103/PhysRevLett.119.170502}
  (\bibinfo{year}{2017}).

\bibitem{doi:10.1126/science.aan0070}
\bibinfo{author}{Kalb, N.} \emph{et~al.}
\newblock \bibinfo{journal}{\bibinfo{title}{Entanglement distillation between
  solid-state quantum network nodes}}.
\newblock {\emph{\JournalTitle{Science}}} \textbf{\bibinfo{volume}{356}},
  \bibinfo{pages}{928--932},
  \doiprefix\url{https://doi.org/10.1126/science.aan0070}
  (\bibinfo{year}{2017}).

\bibitem{li2019experimental}
\bibinfo{author}{Li, Z.-D.} \emph{et~al.}
\newblock \bibinfo{journal}{\bibinfo{title}{Experimental quantum repeater
  without quantum memory}}.
\newblock {\emph{\JournalTitle{Nature Photonics}}}
  \textbf{\bibinfo{volume}{13}}, \bibinfo{pages}{644--648},
  \doiprefix\url{https://doi.org/10.1038/s41566-019-0468-5}
  (\bibinfo{year}{2019}).

\bibitem{Yan_2021}
\bibinfo{author}{Yan, P.-S.}, \bibinfo{author}{Zhou, L.},
  \bibinfo{author}{Zhong, W.} \& \bibinfo{author}{Sheng, Y.-B.}
\newblock \bibinfo{journal}{\bibinfo{title}{A survey on advances of quantum
  repeater}}.
\newblock {\emph{\JournalTitle{{EPL} (Europhysics Letters)}}}
  \textbf{\bibinfo{volume}{136}}, \bibinfo{pages}{14001},
  \doiprefix\url{https://doi.org/10.1209/0295-5075/ac37d0}
  (\bibinfo{year}{2021}).

\bibitem{doi:10.1063/1.523304}
\bibinfo{author}{Misra, B.} \& \bibinfo{author}{Sudarshan, E. C.~G.}
\newblock \bibinfo{journal}{\bibinfo{title}{The {Z}eno’s paradox in quantum
  theory}}.
\newblock {\emph{\JournalTitle{Journal of Mathematical Physics}}}
  \textbf{\bibinfo{volume}{18}}, \bibinfo{pages}{756--763},
  \doiprefix\url{https://doi.org/10.1063/1.523304} (\bibinfo{year}{1977}).

\bibitem{Kofman2000}
\bibinfo{author}{Kofman, A.~G.} \& \bibinfo{author}{Kurizki, G.}
\newblock \bibinfo{journal}{\bibinfo{title}{Acceleration of quantum decay
  processes by frequent observations}}.
\newblock {\emph{\JournalTitle{Nature}}} \textbf{\bibinfo{volume}{405}},
  \bibinfo{pages}{546--550}, \doiprefix\url{https://doi.org/10.1038/35014537}
  (\bibinfo{year}{2000}).

\bibitem{BAYINDIR2018141}
\bibinfo{author}{Bayındır, C.} \& \bibinfo{author}{Ozaydin, F.}
\newblock \bibinfo{journal}{\bibinfo{title}{Freezing optical rogue waves by
  zeno dynamics}}.
\newblock {\emph{\JournalTitle{Optics Communications}}}
  \textbf{\bibinfo{volume}{413}}, \bibinfo{pages}{141--146},
  \doiprefix\url{https://doi.org/10.1016/j.optcom.2017.12.051}
  (\bibinfo{year}{2018}).

\bibitem{BAYINDIR2021127096}
\bibinfo{author}{Bayındır, C.}
\newblock \bibinfo{journal}{\bibinfo{title}{Zeno dynamics of quantum chirps}}.
\newblock {\emph{\JournalTitle{Physics Letters A}}}
  \textbf{\bibinfo{volume}{389}}, \bibinfo{pages}{127096},
  \doiprefix\url{https://doi.org/10.1016/j.physleta.2020.127096}
  (\bibinfo{year}{2021}).

\bibitem{PhysRevA.41.2295}
\bibinfo{author}{Itano, W.~M.}, \bibinfo{author}{Heinzen, D.~J.},
  \bibinfo{author}{Bollinger, J.~J.} \& \bibinfo{author}{Wineland, D.~J.}
\newblock \bibinfo{journal}{\bibinfo{title}{Quantum zeno effect}}.
\newblock {\emph{\JournalTitle{Phys. Rev. A}}} \textbf{\bibinfo{volume}{41}},
  \bibinfo{pages}{2295--2300},
  \doiprefix\url{https://doi.org/10.1103/PhysRevA.41.2295}
  (\bibinfo{year}{1990}).

\bibitem{Schäfer2014}
\bibinfo{author}{Sch{\"a}fer, F.} \emph{et~al.}
\newblock \bibinfo{journal}{\bibinfo{title}{Experimental realization of quantum
  zeno dynamics}}.
\newblock {\emph{\JournalTitle{Nature Communications}}}
  \textbf{\bibinfo{volume}{5}}, \bibinfo{pages}{3194},
  \doiprefix\url{https://doi.org/10.1038/ncomms4194} (\bibinfo{year}{2014}).

\bibitem{PhysRevA.69.012303}
\bibinfo{author}{Beige, A.}
\newblock \bibinfo{journal}{\bibinfo{title}{Ion-trap quantum computing in the
  presence of cooling}}.
\newblock {\emph{\JournalTitle{Phys. Rev. A}}} \textbf{\bibinfo{volume}{69}},
  \bibinfo{pages}{012303},
  \doiprefix\url{https://doi.org/10.1103/PhysRevA.69.012303}
  (\bibinfo{year}{2004}).

\bibitem{PhysRevA.87.032112}
\bibinfo{author}{Zheng, W.} \emph{et~al.}
\newblock \bibinfo{journal}{\bibinfo{title}{Experimental demonstration of the
  quantum zeno effect in nmr with entanglement-based measurements}}.
\newblock {\emph{\JournalTitle{Phys. Rev. A}}} \textbf{\bibinfo{volume}{87}},
  \bibinfo{pages}{032112},
  \doiprefix\url{https://doi.org/10.1103/PhysRevA.87.032112}
  (\bibinfo{year}{2013}).

\bibitem{PhysRevLett.87.040402}
\bibinfo{author}{Fischer, M.~C.}, \bibinfo{author}{Guti\'errez-Medina, B.} \&
  \bibinfo{author}{Raizen, M.~G.}
\newblock \bibinfo{journal}{\bibinfo{title}{Observation of the quantum zeno and
  anti-zeno effects in an unstable system}}.
\newblock {\emph{\JournalTitle{Phys. Rev. Lett.}}}
  \textbf{\bibinfo{volume}{87}}, \bibinfo{pages}{040402},
  \doiprefix\url{https://doi.org/10.1103/PhysRevLett.87.040402}
  (\bibinfo{year}{2001}).

\bibitem{PhysRevLett.101.180402}
\bibinfo{author}{Bernu, J.} \emph{et~al.}
\newblock \bibinfo{journal}{\bibinfo{title}{Freezing coherent field growth in a
  cavity by the quantum zeno effect}}.
\newblock {\emph{\JournalTitle{Phys. Rev. Lett.}}}
  \textbf{\bibinfo{volume}{101}}, \bibinfo{pages}{180402},
  \doiprefix\url{https://doi.org/10.1103/PhysRevLett.101.180402}
  (\bibinfo{year}{2008}).

\bibitem{PhysRevLett.105.213601}
\bibinfo{author}{Raimond, J.~M.} \emph{et~al.}
\newblock \bibinfo{journal}{\bibinfo{title}{Phase space tweezers for tailoring
  cavity fields by quantum zeno dynamics}}.
\newblock {\emph{\JournalTitle{Phys. Rev. Lett.}}}
  \textbf{\bibinfo{volume}{105}}, \bibinfo{pages}{213601},
  \doiprefix\url{https://doi.org/10.1103/10.1103/PhysRevLett.105.213601}
  (\bibinfo{year}{2010}).

\bibitem{PhysRevA.86.032120}
\bibinfo{author}{Raimond, J.~M.} \emph{et~al.}
\newblock \bibinfo{journal}{\bibinfo{title}{Quantum zeno dynamics of a field in
  a cavity}}.
\newblock {\emph{\JournalTitle{Phys. Rev. A}}} \textbf{\bibinfo{volume}{86}},
  \bibinfo{pages}{032120},
  \doiprefix\url{https://doi.org/10.1103/PhysRevA.86.032120}
  (\bibinfo{year}{2012}).

\bibitem{Signoles2014}
\bibinfo{author}{Signoles, A.} \emph{et~al.}
\newblock \bibinfo{journal}{\bibinfo{title}{Confined quantum zeno dynamics of a
  watched atomic arrow}}.
\newblock {\emph{\JournalTitle{Nature Physics}}} \textbf{\bibinfo{volume}{10}},
  \bibinfo{pages}{715--719}, \doiprefix\url{https://doi.org/10.1038/nphys3076}
  (\bibinfo{year}{2014}).

\bibitem{Chen2021NPJ}
\bibinfo{author}{Chen, T.} \emph{et~al.}
\newblock \bibinfo{journal}{\bibinfo{title}{Quantum zeno effects across a
  parity-time symmetry breaking transition in atomic momentum space}}.
\newblock {\emph{\JournalTitle{npj Quantum Information}}}
  \textbf{\bibinfo{volume}{7}}, \bibinfo{pages}{78},
  \doiprefix\url{https://doi.org/10.1038/s41534-021-00417-y}
  (\bibinfo{year}{2021}).

\bibitem{tuncer2019work}
\bibinfo{author}{Tuncer, A.}, \bibinfo{author}{Izadyari, M.},
  \bibinfo{author}{Da{\u{g}}, C.~B.}, \bibinfo{author}{Ozaydin, F.} \&
  \bibinfo{author}{M{\"u}stecapl{\i}o{\u{g}}lu, {\"O}.~E.}
\newblock \bibinfo{journal}{\bibinfo{title}{Work and heat value of bound
  entanglement}}.
\newblock {\emph{\JournalTitle{Quantum Information Processing}}}
  \textbf{\bibinfo{volume}{18}}, \bibinfo{pages}{373},
  \doiprefix\url{https://doi.org/10.1007/s11128-019-2488-y}
  (\bibinfo{year}{2019}).

\bibitem{dag2019temperature}
\bibinfo{author}{Dag, C.~B.}, \bibinfo{author}{Niedenzu, W.},
  \bibinfo{author}{Ozaydin, F.}, \bibinfo{author}{Mustecapl{\i}oglu, O.~E.} \&
  \bibinfo{author}{Kurizki, G.}
\newblock \bibinfo{journal}{\bibinfo{title}{Temperature control in dissipative
  cavities by entangled dimers}}.
\newblock {\emph{\JournalTitle{The Journal of Physical Chemistry C}}}
  \textbf{\bibinfo{volume}{123}}, \bibinfo{pages}{4035--4043},
  \doiprefix\url{https://doi.org/10.1021/acs.jpcc.8b11445}
  (\bibinfo{year}{2019}).

\bibitem{Mukherjee2020}
\bibinfo{author}{Mukherjee, V.}, \bibinfo{author}{Kofman, A.~G.} \&
  \bibinfo{author}{Kurizki, G.}
\newblock \bibinfo{journal}{\bibinfo{title}{Anti-zeno quantum advantage in
  fast-driven heat machines}}.
\newblock {\emph{\JournalTitle{Communications Physics}}}
  \textbf{\bibinfo{volume}{3}}, \bibinfo{pages}{8},
  \doiprefix\url{https://doi.org/10.1038/s42005-019-0272-z}
  (\bibinfo{year}{2020}).

\bibitem{PhysRevA.77.062339}
\bibinfo{author}{Wang, X.-B.}, \bibinfo{author}{You, J.~Q.} \&
  \bibinfo{author}{Nori, F.}
\newblock \bibinfo{journal}{\bibinfo{title}{Quantum entanglement via two-qubit
  quantum zeno dynamics}}.
\newblock {\emph{\JournalTitle{Phys. Rev. A}}} \textbf{\bibinfo{volume}{77}},
  \bibinfo{pages}{062339},
  \doiprefix\url{https://doi.org/10.1103/PhysRevA.77.062339}
  (\bibinfo{year}{2008}).

\bibitem{PhysRevA.105.022439}
\bibinfo{author}{Ozaydin, F.}, \bibinfo{author}{Bayindir, C.},
  \bibinfo{author}{Altintas, A.~A.} \& \bibinfo{author}{Yesilyurt, C.}
\newblock \bibinfo{journal}{\bibinfo{title}{Nonlocal activation of bound
  entanglement via local quantum zeno dynamics}}.
\newblock {\emph{\JournalTitle{Phys. Rev. A}}} \textbf{\bibinfo{volume}{105}},
  \bibinfo{pages}{022439},
  \doiprefix\url{https://doi.org/10.1103/PhysRevA.105.022439}
  (\bibinfo{year}{2022}).

\bibitem{PhysRevLett.82.1056}
\bibinfo{author}{Horodecki, P.}, \bibinfo{author}{Horodecki, M.} \&
  \bibinfo{author}{Horodecki, R.}
\newblock \bibinfo{journal}{\bibinfo{title}{Bound entanglement can be
  activated}}.
\newblock {\emph{\JournalTitle{Phys. Rev. Lett.}}}
  \textbf{\bibinfo{volume}{82}}, \bibinfo{pages}{1056--1059},
  \doiprefix\url{https://doi.org/10.1103/PhysRevLett.82.1056}
  (\bibinfo{year}{1999}).

\bibitem{CHEN2016140}
\bibinfo{author}{Chen, Y.-H.}, \bibinfo{author}{Huang, B.-H.},
  \bibinfo{author}{Song, J.} \& \bibinfo{author}{Xia, Y.}
\newblock \bibinfo{journal}{\bibinfo{title}{Transitionless-based shortcuts for
  the fast and robust generation of w states}}.
\newblock {\emph{\JournalTitle{Optics Communications}}}
  \textbf{\bibinfo{volume}{380}}, \bibinfo{pages}{140--147},
  \doiprefix\url{https://doi.org/10.1016/j.optcom.2016.05.068}
  (\bibinfo{year}{2016}).

\bibitem{doi:10.1126/science.aaa0754}
\bibinfo{author}{Barontini, G.}, \bibinfo{author}{Hohmann, L.},
  \bibinfo{author}{Haas, F.}, \bibinfo{author}{Estève, J.} \&
  \bibinfo{author}{Reichel, J.}
\newblock \bibinfo{journal}{\bibinfo{title}{Deterministic generation of
  multiparticle entanglement by quantum zeno dynamics}}.
\newblock {\emph{\JournalTitle{Science}}} \textbf{\bibinfo{volume}{349}},
  \bibinfo{pages}{1317--1321},
  \doiprefix\url{https://doi.org/10.1126/science.aaa0754}
  (\bibinfo{year}{2015}).

\bibitem{DetExp}
\bibinfo{author}{Yesilyurt, C.} \emph{et~al.}
\newblock \bibinfo{journal}{\bibinfo{title}{Deterministic local doubling of {W}
  states}}.
\newblock {\emph{\JournalTitle{J. Opt. Soc. Am. B}}}
  \textbf{\bibinfo{volume}{33}}, \bibinfo{pages}{2313},
  \doiprefix\url{https://doi.org/10.1364/JOSAB.33.002313}
  (\bibinfo{year}{2016}).

\bibitem{CavityRefs2}
\bibinfo{author}{Zang, X.-P.}, \bibinfo{author}{Yang, M.},
  \bibinfo{author}{Ozaydin, F.}, \bibinfo{author}{Song, W.} \&
  \bibinfo{author}{Cao, Z.-L.}
\newblock \bibinfo{journal}{\bibinfo{title}{Generating multi-atom entangled {W}
  states via light-matter interface based fusion mechanism}}.
\newblock {\emph{\JournalTitle{Sci. Rep.}}} \textbf{\bibinfo{volume}{5}},
  \bibinfo{pages}{16245}, \doiprefix\url{https://doi.org/10.1038/srep16245}
  (\bibinfo{year}{2015}).

\bibitem{CavityRefs21}
\bibinfo{author}{Zang, X.-P.}, \bibinfo{author}{Yang, M.},
  \bibinfo{author}{Ozaydin, F.}, \bibinfo{author}{Song, W.} \&
  \bibinfo{author}{Cao, Z.-L.}
\newblock \bibinfo{journal}{\bibinfo{title}{Deterministic generation of large
  scale atomic {W} states}}.
\newblock {\emph{\JournalTitle{Opt. Exp(11)}}}
  \textbf{\bibinfo{volume}{24(11)}}, \bibinfo{pages}{12293},
  \doiprefix\url{https://doi.org/10.1364/OE.24.012293} (\bibinfo{year}{2015}).

\bibitem{SRep_2020_10_3481}
\bibinfo{author}{Bugu, S.}, \bibinfo{author}{Ozaydin, F.},
  \bibinfo{author}{Ferrus, T.} \& \bibinfo{author}{Kodera, T.}
\newblock \bibinfo{journal}{\bibinfo{title}{Preparing multipartite entangled
  spin qubits via pauli spin blockade}}.
\newblock {\emph{\JournalTitle{Scientific Reports}}}
  \textbf{\bibinfo{volume}{10}}, \bibinfo{pages}{3481},
  \doiprefix\url{https://doi.org/10.1038/s41598-020-60299-6}
  (\bibinfo{year}{2020}).

\bibitem{PRA_2021_103_052421}
\bibinfo{author}{Ozaydin, F.}, \bibinfo{author}{Yesilyurt, C.},
  \bibinfo{author}{Bugu, S.} \& \bibinfo{author}{Koashi, M.}
\newblock \bibinfo{journal}{\bibinfo{title}{Deterministic preparation of $w$
  states via spin-photon interactions}}.
\newblock {\emph{\JournalTitle{Phys. Rev. A}}} \textbf{\bibinfo{volume}{103}},
  \bibinfo{pages}{052421},
  \doiprefix\url{https://doi.org/10.1103/PhysRevA.103.052421}
  (\bibinfo{year}{2021}).

\bibitem{PhysRevA.65.032314}
\bibinfo{author}{Vidal, G.} \& \bibinfo{author}{Werner, R.~F.}
\newblock \bibinfo{journal}{\bibinfo{title}{Computable measure of
  entanglement}}.
\newblock {\emph{\JournalTitle{Phys. Rev. A}}} \textbf{\bibinfo{volume}{65}},
  \bibinfo{pages}{032314},
  \doiprefix\url{https://doi.org/10.1103/PhysRevA.65.032314}
  (\bibinfo{year}{2002}).

\end{thebibliography}

\section*{Acknowledgements}

VB thanks to Onur Kaya for fruitful discussions. FO acknowledges the Personal Research Fund of Tokyo International University.

\section*{Author contributions statement}

V.B. designed the scheme and carried out the theoretical analysis under the guidance of F.O. V.B., F.O. reviewed the manuscript and contributed to the interpretation of the work and the writing of the manuscript.

\section*{Competing interests}

The authors declare no competing interests.
\ \\ \ \\ \ \\

\begin{figure}
	\centering
	\includegraphics[width=0.7\columnwidth]{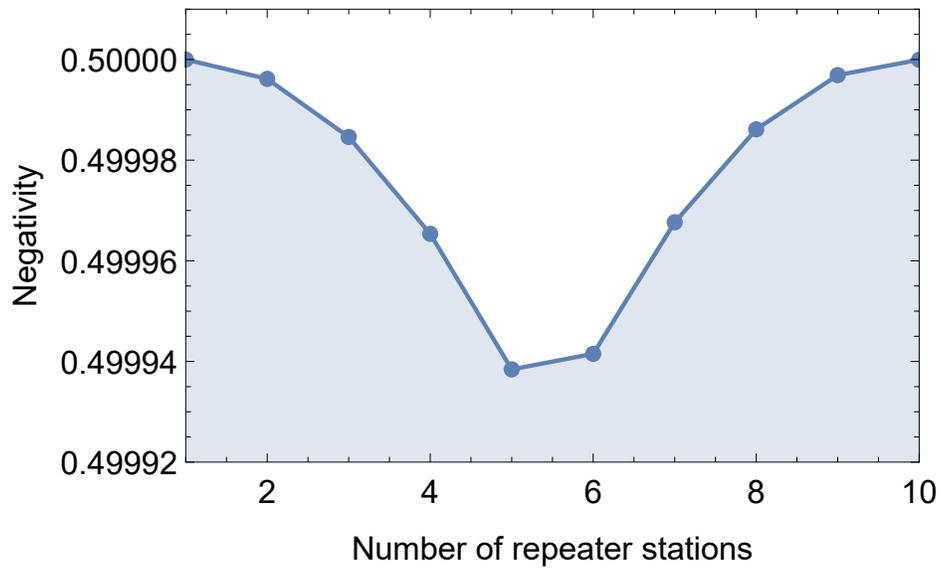}
	\caption{Negativity of the obtained state after ES over a few repeater stations for demonstrating one of the periodic turning points.}
	\label{fig:PlotRepeaterFirst10}
\end{figure}

\end{document}